\documentclass[12pt,preprint]{emulateapj}
\newcommand{\na}{{\it New Astronomy}}

\slugcomment{Submitted to ApJ}

\shorttitle{X-rays in NGC 7027}
\shortauthors{Montez et al.}

\begin{document} 

\title{Dissecting the X-ray Emission in the Young Planetary Nebula NGC 7027}

\author{Rodolfo Montez Jr.}
\affil{Smithsonian Astrophysical Observatory, 60 Garden Street, Cambridge, MA 02138, USA}
\email{rodolfo.montez.jr@gmail.com}

\and 

\author{Joel H. Kastner} 
\affil{Center for Imaging Science and Laboratory for Multiwavelength Astrophysics, Rochester Institute of Technology, 54 Lomb Memorial Drive, Rochester, NY 14623, USA}

\begin{abstract}
We present analysis of a second observation of the young planetary nebula (PN) NGC 7027 by the {\it Chandra} X-ray Observatory. 
This latest 59.2 ks exposure with ACIS-S was acquired approximately 14 years after the initial 18.2 ks ACIS-S observation, and the improved photon statistics allow us to perform a detailed spatial and spectral analysis of the X-ray emission. 
Comparison with multiwavelength imaging of NGC 7027 reveals a strong anti-correlation between extinction across the nebula and the soft band X-ray emission. 
Dissecting the X-ray emission into low- and high-extinction regions results in more robust characterization of the plasma spectral properties.  
We determine that the X-ray emitting plasma has a temperature of $\sim3.6$~MK, is deficient in Fe, and has an X-ray luminosity of $L_{\rm X}\sim7\times10^{31}{\rm ~erg~s}^{-1}$, all of which are generally consistent with the plasma properties found in PN hot bubbles. 
We find no evidence of evolution in the X-ray surface brightness over the 14 year baseline between CXO observations. 
Our analysis underscores the importance of accounting for nebular absorption of the X-ray emission in accurately determining plasma properties for hot bubbles within PNe. 
\end{abstract}


\section{Introduction}

Planetary nebulae (PNe) represent very late stages in the evolution of stars (1 to 8 $M_{\odot}$), when mass loss during the asymptotic giant branch (AGB) phase is swept into shells and ionized by winds and UV from the exposed hot stellar core \citep{1978ApJ...219L.125K}. 
PNe mostly exhibit nonspherical or axisymmetric morphologies, especially  elliptical and bipolar morphologies, with round morphologies appearing less frequently \citep{1995A&A...293..871C,2002ARA&A..40..439B,2009PASP..121..316D}.
Evidence for episodic, apparently spherical mass loss during the AGB is present around some asymmetric nebulae, suggesting a transition from spherical mass loss to the collimated mass loss processes that shape the PN \citep[e.g.,][]{2003MNRAS.340..417C,2004A&A...417..637C}.  
Interacting binary systems have been proposed as the mechanism that produces profoundly asymmetric structures in elliptical and bipolar PNe \citep[e.g.,][]{2000ApJ...538..241S,2008NewA...13..563A}. 
In such a scenario a collimated wind or jet, possibly launched by a disk and/or magnetic process, is responsible for the asymmetry.
A similar process is seen in symbiotic interacting binary systems, R Aqr being a well-studied example. 
R Aqr displays a large jet seen in the infrared and optical, and X-ray monitoring has revealed the proper motion of the jet as it interacts with the nebular material \citep{2001ApJ...563L.151K,2007ApJ...664.1079K}. 

X-ray emission from PNe has been studied extensively by the {\it Chandra} X-ray Observatory and the {\it XMM-Newton} Observatory \citep[e.g.,][]{2002A&A...387L...1G,2005ApJ...635..381M,2006ApJ...653..339G,2008ApJ...672..957K,2013ApJ...767...35R}. 
In particular, the {\it Chandra} Planetary Nebulae Survey \citep[ChanPlaNS;][]{2012AJ....144...58K,2014ApJ...794...99F,2015ApJ...800....8M} has provided insight into the variety of X-ray emission sources within PNe. 
Extended sources of X-ray emission are commonly associated with the so-called PN hot bubble \citep{2008ApJ...672..957K,2012AJ....144...58K,2013ApJ...767...35R,2014ApJ...794...99F}, i.e., the tenuous post-shock region that theory predicts will fill the recently formed nebular cavity \citep{1985A&A...153...79V,2008A&A...489..173S}. 
In the case of the young PN NGC 7027, a {\it Chandra} X-ray observation \citep{2001ApJ...550L.189K} revealed a multi-lobed structure in the extended X-ray emission that seemed closely related to the morphology of the molecular envelope seen in near infrared HI and $H_2$ images \citep{2000ApJ...539..783L,2002A&A...384..603C}. 
This early {\it Chandra} observation led \citet{2003ApJ...583..368S} to consider whether the X-ray emission arose from slower collimated flows launched during the AGB phase of the progenitor star. 
\citet{2008NewA...13..563A} studied a scenario wherein a collimated flow is injected into a spherically symmetric AGB wind, leading to the formation of an asymmetric PN. 
These simulations yield estimates for the luminosities and temperatures of X-ray emission that match the observed X-ray properties for a number of PNe and produce morphological properties similar to those seen in the X-ray emission from NGC 7027. 
These results indicate the influence of an additional process that can also occur in tandem with the typical hot bubble wind interaction process. 

In this article we analyze a second {\it Chandra} X-ray observation of the young PN NGC 7027. 
Our newer, deeper exposure was acquired 14 years after the first exposure and allows for a detailed study of the spatial and spectral distributions of the X-ray emission detected from NGC 7027. 
In \S\ref{obs_sec} we describe the {\it Chandra} observations. 
Our analysis procedures and results are detailed in \S\ref{analysis_sec}, and the results are discussed in comparison with previous observations and with respect to hot bubbles from PNe in \S\ref{discussion_sec}. 
In \S\ref{conclusion_sec} we summarize our conclusions. 

\section{Observations}\label{obs_sec}

NGC 7027 was initially targeted by the {\it Chandra} X-ray Observatory with its back-illuminated ACIS-S detector in 2000 \citep[18.2 ks on 2000-06-01; ObsID 588;][]{2001ApJ...550L.189K}. 
A second ACIS-S observation was acquired nearly fourteen years later (59.2 ks on 2014-03-21; ObsID 15736).  
The increase in exposure time was intended to mitigate the decrease in ACIS sensitivity to soft X-ray photons ($E \lesssim 1 {\rm ~keV}$) \citep{2004SPIE.5165..497M}. 
Caution must be exercised when comparing the two observations due to uncertainty in the changes in instrument sensitivity over time. 

We reprocessed the 2000 and 2014 X-ray observations with CIAO (version 4.6) using the \verb+chandra_repro+ task and calibration database (CALDB version 4.5.9).  
We applied the \verb+EDSER+ option, which uses the energy-dependent sub-pixel event repositioning (SER) algorithm to optimize the spatial resolution of CXO/ACIS-S observations, yielding an effective PSF FWHM of $\sim0\farcs3$ \citep[see][]{2003ApJ...590..586L, 2004ApJ...610.1204L}.  
 
We also include in our analysis a narrow-band $H\alpha$ (F656N; $\lambda_{\rm eff} = 656.3 {\rm ~nm}$) filter image acquired on 2008-08-13 by the Hubble Space Telescope (HST) Wide Field Planetary Camera 2 (WFPC2). 
Finally, we adapted the extinction map of NGC 7027 presented in \citet{1988A&A...200L..21W}, which is derived from mid- to late-1980's observations with the Very Large Array (VLA) and the 2.5-m Isaac Newton Telescope (INT).

\begin{figure*}
\centering
\includegraphics[width=6in]{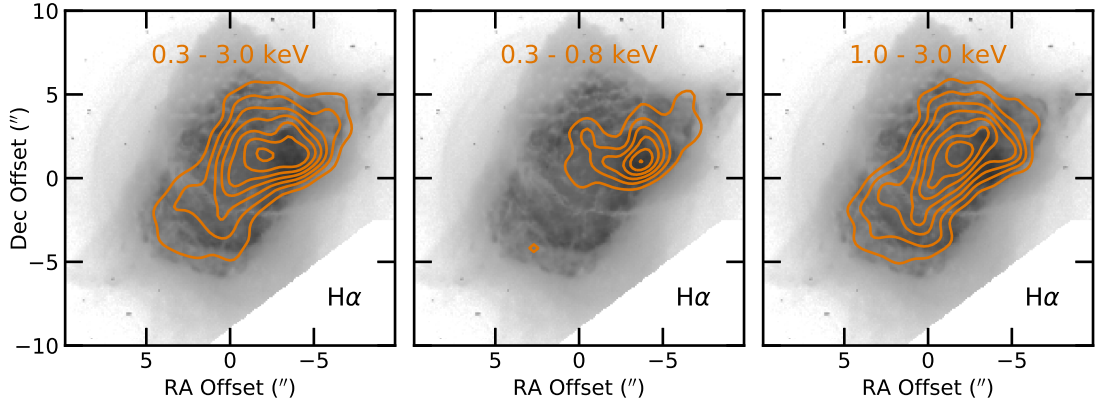}
\caption{Contours of smoothed X-ray emission imaged by {\it Chandra} in 2014 overlaid on the log-scaled H$\alpha$ HST image of NGC 7027. X-ray emission has been smoothed with a Gaussian filter with a FWHM of $\sim1^{\prime\prime}$. X-ray contours are shown for the following three energy-filters: 0.3-3.0 keV (left panel), 0.3-0.8 keV (middle panel), 1.0-3.0 keV (right panel).  \label{fig1}}
\end{figure*}

\begin{figure*}
\centering
\includegraphics[width=6in]{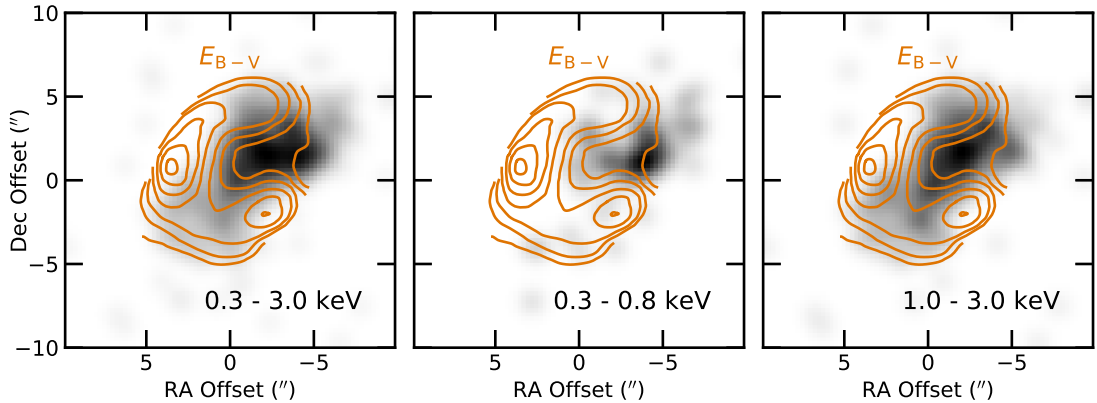}
\caption{Extinction map ($E_{\rm B-V}$) from \citet{1988A&A...200L..21W} is overlaid upon the smoothed X-ray emission from the three energy-filtered images in Figure~\ref{fig1}. The contours span a $E_{\rm B-V}$ range of 0.625 to 1.5 mag with increments of 0.125 mag. \label{fig2}}
\end{figure*}

\section{Analysis and Results}\label{analysis_sec}

\subsection{Energy-filtered X-ray Images} 

We created energy-filtered X-ray images of NGC 7027 from the 2000 and 2014 observations. 
The ``full'' energy band includes photon energies between 0.3 and 3.0 keV, the ``soft'' band from 0.3 to 0.8 keV, and the ``hard'' band from 1.0 to 3.0 keV. 
The full energy band spans the complete range of X-ray emission detected from NGC 7027. 
The soft band includes potential emission lines from C, N, O, Fe and continuum, while the hard band includes potential emission lines from Ne, Mg, and Fe.
Note that any Ne IX emission line present ($\sim 0.9$~keV) is excluded from both ``soft'' and ``hard'' bands. 
In Figure~\ref{fig1}, the contours of the smoothed energy-filtered images are overlaid upon the HST H$\alpha$ images. 
Smoothing was performed on the X-ray images by convolution of the image with a normalized Gaussian kernel that has a FWHM of $\sim 1^{\prime\prime}$.

In the full-band images, most of the surface brightness of X-ray emission is contained within a $8^{\prime\prime}\times14^{\prime\prime}$ region aligned with and confined within the optical emission of the nebula.
However, the soft-band and hard-band images display distinctive spatial distributions of X-ray surface brightness. 
In particular, the soft-band image displays a strong surface brightness asymmetry that coincides with a bright spot seen in the $H\alpha$ image. 
This asymmetric emission in the soft-band image is restricted to a compact region ($6^{\prime\prime}\times7^{\prime\prime}$) northwest of the central star position. 
By comparison, the hard-band image morphology appears as an axisymmetric surface brightness structure. 
This axisymmetric emission is aligned with the general axis of the optical nebula and appears brighter to the northwest of the central star position. 
This comparison of hard- and soft-band X-ray images hence suggests that the departure from axisymmetry seen in the full-band images is due mostly to the strong soft-band asymmetry. 

\subsection{X-ray Surface Brightness Relation to Extinction}\label{xsbext}

The surface brightness variations in the soft-band and hard-band images (Figure~\ref{fig1}) were also noted in the 2000 observation presented by \citet{2001ApJ...550L.189K}. 
By comparing with extinction maps derived from near infrared imaging observations, \citet{2002ApJ...581.1225K} suggested that the X-ray surface brightness was anti-correlated with extinction. 
Optical images of NGC 7027 also show surface brightness variations that appear correlated with the extinction across the nebula.  
On the other hand, the infrared and radio images of the nebula \citep{2000ApJ...539..783L,2002A&A...384..603C,2016ApJ...833..115L} appear more symmetric.
This multiwavelength behavior is expected if extinction is the origin since both optical photons and soft-band X-rays ($E \lesssim 1.0 {\rm ~keV}$) are more susceptible to absorption than infrared and radio emission.

\citet{1988A&A...200L..21W} used 2 cm radio continuum and optical H$\beta$ emission to derive a spatially-resolved extinction map of the nebula. 
Because the radio continuum and H$\beta$ line emission each have the same density dependence ($\propto n_{e}^{2}$) their ratio only has a small dependence on the electron temperature ($\propto T_{e}^{0.53}$). 
For constant $T_{e}$ throughout the nebula \citep[$\sim 14000 {\rm ~K}$;][]{1988A&A...200L..21W}, deviations from the expected H$\beta$ line emission derived from the radio emission and the measured H$\beta$ line emission are likely entirely due to extinction. 
In Figure~\ref{fig2}, we compare the extinction map from \citet{1988A&A...200L..21W} with the smoothed energy-filtered X-ray images.
These comparisons demonstrate how strongly anti-correlated the X-ray surface brightness appears with respect to the extinction across the nebula. 
The soft-band emission shows the strongest anti-correlation, with the soft photons only emerging from regions where the extinction drops below $\sim0.75$ mag. 

\subsection{Spectral Analysis}\label{spec_sec}

\subsubsection{Extinction-based Extraction Regions} 

We now consider the low-resolution ACIS-S CCD X-ray spectroscopy to further study the role of extinction on the X-ray surface brightness. 
We identified three regions for spectral extraction based on the extinction map presented in Figure~\ref{fig2}. 
One region encompasses all the X-ray emission from NGC 7027 while the other two correspond to regions of high and low extinction (see inset of Figure~\ref{fig3}). 
For the background spectrum we identified a single, source-free region. 
We extracted the source and background spectra for each region using the \verb+specextract+ CIAO task.
The resulting spectra are presented in Figure~\ref{fig3}.

\subsubsection{Modeling the X-ray Spectra}\label{spectralanalysis}

To model the X-ray spectra, we adopt an optically-thin thermal plasma model based on the Astrophysical Plasma Emission Code  \citep[APEC:][]{2012ApJ...756..128F,2001ApJ...556L..91S} with variable abundances and intervening absorption described by the multiplicative absorption model {\it wabs} \citep{1983ApJ...270..119M}.
The elemental abundances of C, N, O, Ne, and Fe are particularly important over the energy range of photons detected from NGC 7027 and can influence the plasma properties.
We used XSPEC \citep[version 12.7.1;][]{1996ASPC..101...17A} to perform the least-squares minimization between the model and background-subtracted spectra and derive best-fit model parameters. 
The XSPEC {\verb+error+} command was used to derive the 90\% confidence intervals of best-fit model parameters.

\begin{deluxetable*}{lccrrrrrrcrl}
\tabletypesize{\footnotesize}
\tablecaption{Best-fit Model Parameters for the X-ray Spectra of NGC 7027\label{spectralfits}}
\tablewidth{0pt}
\tablehead{\colhead{Region} & \colhead{$N_H$} & \colhead{$T_{\rm X}$} & \colhead{norm.} & \colhead{Fe abund$^{\dagger}$} & \colhead{$F_{\rm X}^{\dagger\dagger}$} & \colhead{$L_{\rm X}^{\dagger\dagger\dagger}$} \\
 & ($10^{22} {\rm ~cm}^{-2}$)  &   (MK) & ($\times 10^{-4}$) & & (c.g.s.) & ($10^{31} {\rm ~erg~s}^{-1}$) }
\startdata

\sidehead{{\it $T_{\rm X}$ and Fe fit simultaneously:}}

Low-$E_{\rm B-V}$ & 0.58$_{-0.19}^{+0.15}$ & 3.6$_{-0.6}^{+0.9}$ & 0.95$_{-0.52}^{+0.97}$ & $0.28^{+0.24}_{-0.16}$ & 1.65$_{-0.90}^{+1.68}$ & $1.53_{-0.87}^{+1.58}$  \\
High-$E_{\rm B-V}$ & 1.01$_{-0.22}^{+0.18}$ & -- & 3.61$_{-1.98}^{+3.80}$ & -- & 6.25$_{-3.43}^{+6.58}$ & $5.79_{-3.33}^{+6.18}$ \\

\sidehead{{\it Fe fixed, remaining parameters fit individually:}}

Full Nebula & 0.80$_{-0.17}^{+0.12}$ & 3.5$_{-0.4}^{+1.0}$ & 4.00$_{-2.20}^{+2.74}$ & 0.28 & 6.95$_{-3.82}^{+4.75}$ & $6.44_{-3.71}^{+4.54}$  \\
Low-$E_{\rm B-V}$ & 0.49$_{-0.17}^{+0.22}$ & 4.0$_{-0.9}^{+1.2}$ & 0.63$_{-0.33}^{+1.20}$ & 0.28 & 1.03$_{-0.55}^{+1.95}$ & $0.95_{-0.53}^{+1.81}$  \\
High-$E_{\rm B-V}$ & 1.05$_{-0.20}^{+0.20}$ & 3.4$_{-0.7}^{+1.0}$ & 4.46$_{-2.68}^{+9.21}$ & 0.28 & 7.78$_{-4.68}^{+16.07}$ & $7.21_{-4.51}^{+14.94}$  \\ 

\enddata
\tablecomments{ \\
$\dagger$ - The Fe abundance is reported with respect to the solar values \citep{1989GeCoA..53..197A}. \\
$\dagger\dagger$ - The unabsorbed flux intrinsic to the source after correcting for intervening absorption is reported for the 0.3 - 3.0 keV energy range and in c.g.s. units of $10^{-13}{\rm ~erg~cm}^{-2}{\rm ~s}^{-1}$.\\
$\dagger\dagger\dagger$ - The X-ray luminosity for the 0.3 - 3.0 keV energy range assuming a distance and error of $0.88\pm0.15$ kpc \citep{1989ApJ...336..294M}.}
\end{deluxetable*}

\begin{figure}
\includegraphics[width=3.4in]{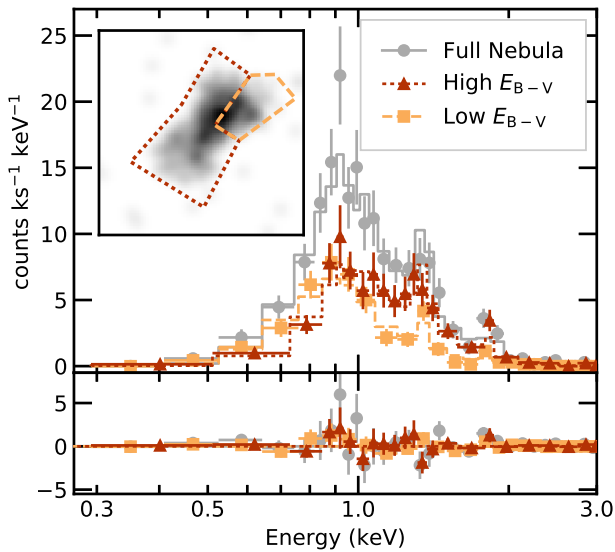}
\centering
\caption{\label{fig3} X-ray spectra of NGC 7027. Circles indicate the X-ray spectrum extracted from the entire region, while triangles and squares indicate the X-ray spectra extracted from the regions of high and low extinction, respectively. Best-fit models for the entire, high-extinction, and low-extinction region are indicated by solid, dotted, and dashed histograms, respectively. The inset shows the high- and low-extraction regions (dotted and dashed lines, respectively) overlaid upon the smoothed full-band image.  }
\end{figure}

We found that, by linking plasma conditions across both the high- and low-extinction regions, the low-count and low-resolution CCD spectra are sufficient to constrain the model parameters, whereas individually the spectra cannot provide sufficient constraints. 
After testing various modeling approaches, we determined the optimal approach is to assume that the plasma temperature and abundances are identical in the low- and high-extinction regions and to simultaneously fit the spectra for the two regions. 
Adopting this approach, we find that the only elemental abundance with statistically significant variations from solar values is Fe. 
As a result, the six free parameters are the (single) plasma temperature,  abundance of Fe relative to solar, and the absorbing columns and plasma emission measures in the low- and high-extinction regions. 
Under these assumptions, all the model parameters can be constrained by simultaneously fitting the two low-resolution spectra from the two extinction-based regions (Figure~\ref{fig3}). 

For comparison, we considered two additional spectral-modeling approaches. 
In one approach we relaxed the assumption of a single plasma temperature in the low- and high-extinction regions. 
In the other approach we fit the X-ray spectrum for a single region that encompasses all of the X-ray emission. 
The former approach is equivalent to fitting each extinction-based spectral extraction region separately, and hence suffers from a significant reduction in the number of counts being used to constrain the model parameters.\footnote{In one of these we also attempted to fit a single column density in the two extinction regions but could not constrain the plasma properties of the low-extinction region, suggesting that such a model is unphysical.}
The latter approach is similar to that used in previous studies of the X-ray emission from NGC 7027 and assumes a single temperature and single extinction value throughout the X-ray emitting region.
With these approaches, we find we can only constrain the model parameters by reducing the number of free parameters, in particular, by fixing the Fe abundance to the value determined from the simultaneous fit (see Table~\ref{spectralfits}). 

\begin{figure}
 \centering
 \includegraphics[width=3.4in]{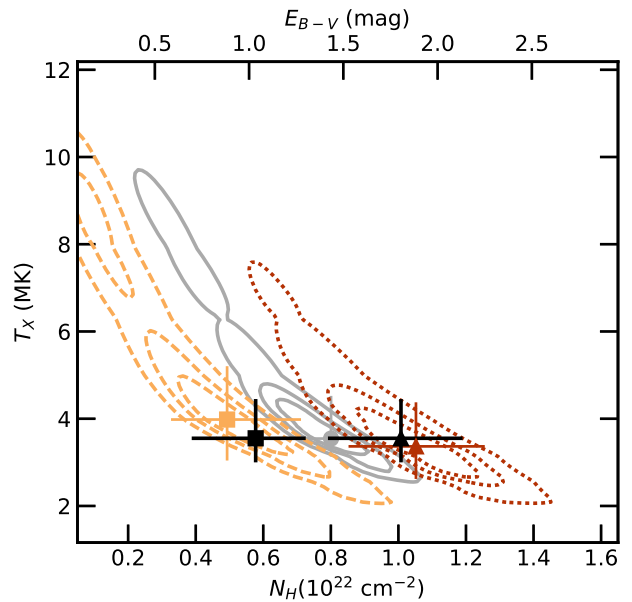}
\caption{\label{fig4} Best-fit confidence ranges for the plasma temperature, $T_{\rm X}$, and column density, $N_H$. The confidence ranges and best-fit values are indicated by distinct symbols and lines as given in Figure~\ref{fig3} for the various spectral extraction regions described in the text. The two black symbols indicate the result from a simultaneous fit to the high- and low-extinction regions assuming a single temperature plasma.  }
\end{figure}

In Table~\ref{spectralfits} we report the resulting best-fit parameters for the foregoing spectral modeling approaches, along with the intrinsic source fluxes and luminosities derived from the spectral models after correcting for absorption.
For all of these individual fits, we used the XSPEC {\verb+steppar+} command to obtain the confidence contours for the best-fit parameter values of the plasma temperature and column density (Figure~\ref{fig4}). 

\subsection{Comparing the 2000 and 2014 Observations}

The potential evolution of the X-ray surface brightness between the 2000 and 2014 observations could provide valuable information concerning nebular kinematics and shock evolution. 
However, direct comparison is hampered by the degradation of the soft X-ray ($<1$ keV) sensitivity of the ACIS-S detector since launch due to a contaminant build-up on the detector optical blocking filter \citep{2004SPIE.5165..497M}. 
Specifically, for the two NGC 7027 observations over the 0.3-3.0 keV energy range, we find that the count rate dropped from $\sim15$~cnt~ks$^{-1}$ in 2000 to $\sim9.4$~cnt~ks$^{-1}$ in 2014. 
The drop is more pronounced in the soft energy range (0.3-0.8 keV), where the count rates dropped from $\sim3.0$~cnt~ks$^{-1}$ in 2000 to $\sim1.2$~cnt~ks$^{-1}$ in 2014. 
Such a drop in count rate is consistent with the expected decreased sensitivity to soft X-ray photons. 
In the following, we attempt to mitigate this loss of sensitivity by focusing on the hard-band images ($> 1$~keV), where only slight changes to sensitivity have occurred. 

\begin{figure*}[ht]
\centering
\includegraphics[width=6in]{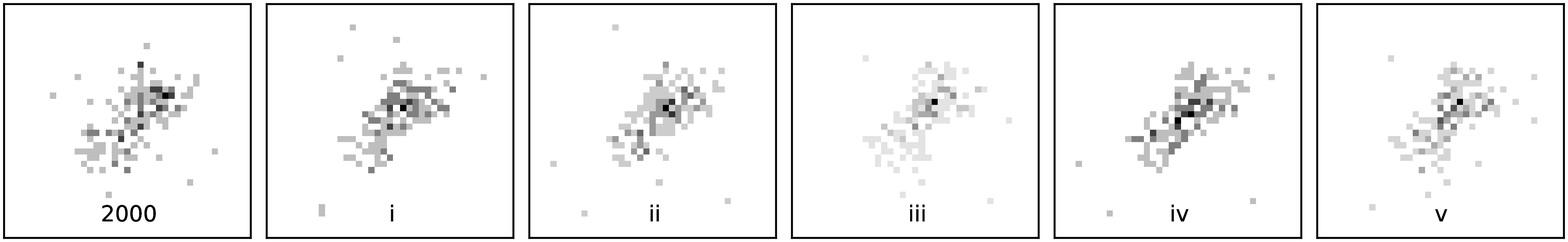}
\caption{Spatial distribution of the hard-band X-ray emission. The left-most panel shows the hard-band (1.0-3.0 keV) image of the full 2000 observation. 
The remaining panels show randomly-sampled hard-band images of the 2014 observation constructed by limiting the images to the same number of photons (140)  detected in the entire 2000 exposure. \label{samples}}
\end{figure*}

The longer exposure time of the 2014 observation led to better photon statistics in that (second) observation. 
This, in turn, allowed us to test the impact of Poisson fluctuations on the hard-band X-ray surface brightness. 
We accomplish this by randomly sampling $N$ hard photons from the 2014 observation, where $N$ is set equal to the number of hard photons detected in the entire 2000 exposure ($N = 140$). 
In Figure~\ref{samples} we present five 2014 hard-band images generated from this random sampling test alongside the 2000 hard-band image. 
The comparison suggests that Poisson fluctuations likely dominate the details in the structures seen in the 2000 observation. 
In particular, the protrusions towards the northwest, which are prominent in the 2000 epoch hard-band images  \citep{2001ApJ...550L.189K}, are not present in the more recent observation (Figure~\ref{fig2}). 
Based on this analysis (Figure~\ref{samples}) and barring any unknown instrumental change in the spatial resolution, we conclude there is no evidence for expansion or proper motion of the X-ray emission on scales $>1^{\prime\prime}$, and that any attempt to quantify expansion on smaller scales is highly uncertain due to the Poisson fluctuations of the low-count observations. 

\section{Discussion}\label{discussion_sec}

\subsection{Spatially-varying Absorption across the Nebula}

The spatial anti-correlation between the extinction map and soft-band X-ray emission  (Figure~\ref{fig2}) combined with the results of our spectral fitting (Figures~\ref{fig3} \& ~\ref{fig4}) demonstrate that the detailed X-ray emission surface brightness distribution --- and its interpretation --- is sensitive to the spatial variations in extinction. 
To quantify this spatially-varying extinction we have identified and analyzed two distinct regions of the X-ray emission detected from NGC 7027 (\S\ref{spec_sec}). 
We found best-fit values of $N_{\rm H}=6\times10^{21} {\rm ~cm}^{-2}$ and $N_{\rm H} = 10\times10^{21} {\rm ~cm}^{-2}$ for the absorbing columns towards the low- and high-extinction regions, respectively. 
These absorption values represent the averages in the regions studied, whereas in reality the absorption varies continuously across the nebula, as indicated by the extinction maps from \citet{1988A&A...200L..21W}. 
Adopting the relationships $N_{\rm H} = 1.79 \times 10^{21} {\rm ~cm}^{-2} {\rm ~mag}^{-1} A_{\rm V}$ \citep{1995A&A...293..889P} and $A_{\rm V} = R_{\rm V} E_{\rm B-V}$ with $R_{\rm V} = 3.1$, we can compare our column densities to the values provided by the \citet{1988A&A...200L..21W} extinction map. 
Our average extinction value for the high extinction region is slightly higher than indicated by the extinction map, while the value for the low extinction region is consistent with the extinction map. 

The slight discrepancy between the $E_{\rm B-V}$ value obtained from the extinction map and the value obtained from the high-extinction region X-ray spectrum is consistent with the suggestion by \citet{1988A&A...200L..21W} that extinction is caused partly by dust associated with the nebula. 
Indeed, high-resolution imaging of dust and molecular emission of NGC 7027 indicate that the molecules and dust in the nebula are likely responsible for its pattern of enhanced extinction \citep{2000ApJ...539..783L,2002A&A...384..603C,2016ApJ...833..115L}. 
Because some dust is associated with the nebula, the values provided by the extinction map are actually lower limits, and our spectroscopically-determined average absorption values provide a better estimate of the degree of extinction caused by the dust in the nebula. 
Based on the general distribution of the X-ray emission in the high extinction region with respect to the extinction map contours (see Figure~\ref{fig2}), the spectral fit suggests the $E_{\rm B-V}$ map underestimates the extinction by $\sim$1 mag. 
This local source of absorption primarily affects the soft-band X-ray photons, suggesting that such X-ray irradiation should have an effect on the ionization and dissociation of the absorbing material. 

\subsection{Plasma Properties and Previous Studies} 

In all spectral fitting methods we considered, the results for plasma temperature are consistent with a roughly isothermal X-ray emitting region (see Table~\ref{spectralfits}). 
The best fit value is $T_{\rm X}\sim3.6$ MK with a 90\% confidence range of 3-4.5 MK. 
The ratio of the emission measures in the two regions studied is consistent with the relative sizes of the regions (see inset of Figure~\ref{fig3}), suggesting the plasma density is also constant, on average, throughout the X-ray emitting region. 
The inferred density of the X-ray emitting plasma, adopting the normalization of the full nebular fit (Table~\ref{spectralfits}) and assuming a fully-ionized plasma with a unity filling factor, is $n_{e} = 120 {\rm ~cm}^{-3}$ with a 90\% confidence interval of $50-200{\rm ~cm}^{-3}$. 
Together, these parameters suggest the pressure in the hot gas is $5\times10^{-9} {\rm ~dynes~cm}^{-2}$ with a 90\% confidence interval of $2-8\times10^{-9} {\rm ~dynes~cm}^{-2}$. 
From the simultaneous fit to the low- and high-extinction regions, the total $L_{\rm X}$ is $7\times10^{31} {\rm ~erg~s}^{-1}$ with a confidence range of $(3-15)\times10^{31} {\rm ~erg~s}^{-1}$.
The upper end of our $L_{\rm X}$ confidence range is consistent with the X-ray luminosity reported in \citet{2001ApJ...550L.189K}, especially given that their adopting a single, uniform value for X-ray absorption should bias $L_{\rm X}$ toward larger values.

Based on the early 2000 Chandra observations of NGC 7027, \citet{2001ApJ...550L.189K} reported $T_{\rm X}\sim3$~MK while \citet{2003ApJ...589..439M} reported temperatures up to 10 MK. 
The results of our analysis are more consistent with those of \citet{2001ApJ...550L.189K}. 
The key difference between our study and that of \citet{2003ApJ...589..439M} is the different abundances found for the X-ray emitting plasma. 
The plasma model adopted by \citet{2003ApJ...589..439M} required modified abundances of multiple elements, with strong enhancements of O and Mg, whereas we only needed to let Fe vary to achieve a satisfactory fit to the X-ray spectrum. 
At the energy resolution of ACIS-S imaging-spectroscopy, the X-ray emitting plasma of NGC 7027 is a blend of emission lines and continuum. 
Depending on the plasma temperature, emission lines from lighter elements like N, O, and Ne can blend with a forrest of Fe XVII lines in the 0.7 to 1.0 keV energy range. 
The potential confusion between the N, O, Ne and Fe emission lines is evident in the fit behavior depicted in Figure~\ref{fig4}. 
Emission lines of Ne IX and Fe XVII arise from plasma spanning a similar range of temperatures (1-10 MK), however, in the higher temperature range ($T_{\rm X}> 4$~MK) the unresolved Fe XVII lines start to contribute more flux in the 0.7-0.95 keV energy range. 
As a result, Fe emission lines (Fe XVII and, possibly, Fe XIX) in the model start to dominate the fit statistic and the spectral fit tends towards higher plasma temperatures.  

This ambiguity in the model due to unresolved Fe lines is also clear in the analysis of hot bubble plasmas reported in \citet{2006ApJ...639..185G}. 
Specifically, when fitting the X-ray spectrum of NGC 7027 with a series of fixed plasma temperatures and solar Fe abundance, the model-derived Ne abundance takes on negligible values at the highest temperature ($T_{\rm X} \sim 10$~MK). 
Additionally, the analysis in \citet{2006ApJ...639..185G} demonstrates how the reduction of O VIII emissivity at higher temperatures leads to an increase in the O abundance required to account for the emission near 0.6 keV.
In our analysis, we find that this ambiguity in spectral fitting is mitigated by separating the extinction regions and leaving the Fe abundance as a free parameter. 
The resulting decrease in Fe abundance then explains the discrepancies between our results and those of \citet{2003ApJ...589..439M}, and supports the conclusions of \citet{2006ApJ...639..185G} as well as analysis of the high-resolution X-ray spectrum of BD+30$^{\circ}$3639 \citep{2009ApJ...690..440Y} both of which indicate that hot bubbles of PNe are depleted in Fe.
Indeed, \citet{1990MNRAS.244..294M} reports that the Fe abundance is 45 times less than solar in the $\sim10^{4} {\rm ~K}$ plasma within NGC 7027, consistent with the trend of large (optically measured) Fe depletions in PNe generally \citep{2009ApJ...694.1335D,2014ApJ...784..173D,2016MNRAS.456.3855D}. 
Such a large Fe depletion in the X-ray emitting plasma would make Fe negligible in our X-ray spectral model. 
Without the advantage of high resolution X-ray spectroscopy, such as that obtained for the planetary nebula BD+30$^{\circ}$3639 \citep{2009ApJ...690..440Y}, the extent of the sub-solar Fe abundance in the X-ray emitting plasma of NGC 7027 will remain uncertain and, in turn, we are limited in the conclusions that can be made regarding the origin of Fe in the X-ray emitting plasma.

\subsection{NGC 7027 in the Hot Bubble Context}

The chemical composition of the X-ray emitting gas plays a vital role in our understanding of the formation and evolution of hot bubbles in planetary nebulae. 
Some studies of X-ray spectra find that the hot bubble chemical abundances are more consistent with the stellar wind abundances \citep[e.g.,][]{2001ApJ...553L..69C,2009ApJ...690..440Y}, indicating that the stellar wind supplies a majority of the hot bubble material. 
However, other studies have argued that the nebula is the origin of the hot bubble material \citep[e.g.,][]{2008ApJ...681..333G}. 
Using 1D hydrodynamics simulations that treat the energy transfer from heat conduction across the hot bubble-nebular interface, \citet{2008A&A...489..173S} argued that nebular material is ``evaporated'' into the hot bubble and therefore over time the nebular mass should dominate over the mass provided by the fast wind to the X-ray emitting plasma. 
The timescale for this process depends on the initial mass of the progenitor, where more massive progenitors transition to nebular-dominated compositions faster than lower mass progenitors. 
Evidence suggests a massive progenitor for the central star of NGC 7027 \citep[$>3M_{\odot}$;][]{1988A&A...200L..21W}, such that even at the relatively young dynamical age of $\sim600$ years \citep{1989ApJ...336..294M}, the X-ray emitting material would have already transitioned to the nebular-dominated composition. 
Indeed, we find the Fe abundance values indicated by the X-ray spectral fits to be consistent with the (depleted) Fe abundances measured for the optical nebula \citep{1990MNRAS.244..294M}, however, the chemical abundances of the stellar wind are unknown. 

If heat conduction plays a dominate role in the characteristics of the extended X-ray emission, then the temperature depends on the wind power and bubble size \citep{2008A&A...489..173S}.
Assuming the central star properties ($M_{*} \sim 0.7 M_{\odot}$ and $R_{*} \sim 0.17 R_{\odot}$) determined by \citet{1990MNRAS.244..294M}, the escape velocity, $v_{\rm esc}$, at the surface of the central star suggests a wind velocity of at least $1200 {\rm ~km~s}^{-1}$.
For a hot bubble radius equal to $\sim 0.026 {\rm ~pc}$, this escape velocity limit combined with the plasma temperature estimate from \citet{2008A&A...489..173S}, their Eq. (14), suggests that the stellar mass loss rate must be less than $5\times10^{-7} ~M_{\odot} {\rm ~yr}^{-1}$, which is consistent with mass loss rates measured from other central stars.
However, no stellar wind from the central star of NGC 7027 has yet been detected. 
If the stellar wind is slower, then the best-fit plasma temperature, which is one of the hottest measured from a PN, is difficult to reconcile with the \citet{2008A&A...489..173S} heat conduction models. 
The X-ray properties and hot bubble radii predicted by \citet{2014MNRAS.443.3486T,2016MNRAS.463.4438T} for their heat conduction model with the highest assumed initial mass -- $2.5~M_{\odot}$, which is less than that deduced for the progenitor star of NGC 7027 \citep[$>3M_{\odot}$;][]{1988A&A...200L..21W} -- are consistent with those of NGC 7027 and the model suggests $L_{\rm wind}\sim10^{0.5}~L_{\odot}$.
Direct measurement of the properties of the stellar wind of the central star of NGC 7027 would help establish the thermal energy provided by the wind so as to constrain role of heat conduction in the X-ray emission from this energetic nebula.

In the case of a strong shock, the plasma temperature provides a diagnostic of the shock velocity, since
\begin{equation}
T_{\rm shock} = \frac{3}{16} \frac{\mu m_{H}}{k} v_{\rm wind}^2 \sim 1.4 \times10^{7}  {\rm ~K} \left( \frac{v_{\rm wind}}{1000 {\rm ~km~s}^{-1}} \right)^{2} ,
\end{equation}
where $\mu$ is the mean molecular weight, which we assume is 0.6 for a fully ionized plasma with solar abundances. 
If we assume $v_{\rm wind} = v_{\rm esc}\sim 1200 {\rm ~km~s}^{-1}$, then $T_{\rm shock}\sim20 {\rm ~MK}$, a factor of $\sim6$ higher then measured $T_{\rm X}$. 
Alternatively, the best-fit plasma temperature implies $v_{\rm wind}\sim 500 {\rm ~km~s}^{-1}$, which is inconsistent with the speed imposed by the central star escape velocity but perhaps more consistent with an outflow such as that seen in the symbiotic system R Aqr \citep{2001ApJ...563L.151K,2007ApJ...664.1079K}. 
Collimated outflows were indicated as the origin of the high-velocity ionized gas (Br$\gamma$) in NGC 7027 \citep{2002A&A...384..603C}. 
 \citet{2002A&A...384..603C} pointed out the close morphological correspondence between high-velocity Br$\gamma$ and X-ray emission, but noted that the outflows detected in Br$\gamma$ could only reach X-ray emitting temperatures if oriented nearly in the plane of the sky. 
In the case of R Aqr, the proper motion of the X-ray emitting regions provides an independent estimate of the outflow velocity \citep{2001ApJ...563L.151K,2007ApJ...664.1079K}. 
For a distance of 0.88 kpc \citep{1989ApJ...336..294M}, the apparent lack of proper motion of the X-ray emission from NGC 7027 constrains the velocity of any putative collimated outflows to be $<300 {\rm ~km~s}^{-1}$, unless the nebular material decelerates (or terminates) the X-ray emitting plasma.  
Hence, the limits on the plane-of-sky expansion of the X-ray-emitting region placed by our second-epoch Chandra observations makes the strong shock scenario less appealing than heat conduction, as a means to explain the measured plasma temperature. 

\section{Conclusions}\label{conclusion_sec}

An early {\it Chandra} observation of NGC 7027 raised a number of questions about the process responsible for the X-ray emission detected from the young planetary nebula \citep{2001ApJ...550L.189K,2003ApJ...583..368S,2003ApJ...589..439M,2006ApJ...639..185G}. 
Our analysis of a second deeper observation acquired 14 years later addresses many of these questions. 
An essential aspect of this analysis is the identification of two distinct regions based on comparisons with a high-resolution extinction map of NGC 7027 \citep{1988A&A...200L..21W}. 
The soft band X-ray emission is anti-correlated with the extinction map, suggesting that the soft X-ray photons are heavily absorbed in the southeast regions of the nebula relative to its northwest regions. 
Dissecting the X-ray emission by using low- and high-extinction regions derived from the extinction map allows us to constrain the global plasma properties (in particular, $T_{\rm X}$ and Fe abundance) and the average column densities in each region. 
The derived column densities are consistent with the extinction map and further indicate that dust within the nebula is partially responsible for absorbing the soft X-ray emission. 
Our analysis also resolves an ambiguity in previous studies with respect to the Fe content of PN hot bubbles. 
We determined that the X-ray-emitting plasma in NGC 7027 is deficient in Fe, which supports results presented in \citet{2006ApJ...639..185G}. 
The Fe content and the plasma temperature are generally consistent with the Fe abundances and plasma temperatures determined for PN hot bubbles \citep[e.g.,][]{2008ApJ...672..957K,2009ApJ...690..440Y}.
Based on these properties, there are two processes that could explain the X-ray emission from NGC 7027. 
Namely, a hot bubble is formed by an undetected fast ($> 1200 {\rm ~km~s}^{-1}$) stellar wind, which is then subject to heat conduction processes such as those described by \citet{2008A&A...489..173S,2014MNRAS.443.3486T,2016MNRAS.463.4438T}. 
Rather than originating from a fast, quasi-spherical stellar wind, the X-ray-generating shocks could be the result of collimated and, possibly, variable speed outflows \citep{2008NewA...13..563A} such as those seen from jets launched by symbiotic systems.
These two scenarios are not mutually exclusive and may each have a role in the production of X-ray emission from NGC 7027. 
The properties of the central star, such as its wind and chemical composition, are key missing ingredients that will help us better understand the X-ray emission from NGC 7027.

\acknowledgments
{\it Facilities:} \facility{HST (WFPC2)}, \facility{CXO (ACIS)}.

Support for this work was provided by the National Aeronautics and Space Administration through Chandra Award Number GO4-15022X issued by the Chandra X-ray Center, which is operated by the Smithsonian Astrophysical Observatory for and on behalf of the National Aeronautics Space Administration under contract NAS8-03060.
The scientific results reported in this article are based to a significant degree on observations made by the Chandra X-ray Observatory, data obtained from the Chandra Data Archive, and observations made by the Chandra X-ray Observatory and published previously in cited articles.
The scientific results are also based on observations made with the NASA/ESA Hubble Space Telescope, and obtained from the Hubble Legacy Archive, which is a collaboration between the Space Telescope Science Institute (STScI/NASA), the Space Telescope European Coordinating Facility (ST-ECF/ESA) and the Canadian Astronomy Data Centre (CADC/NRC/CSA).
This research has made use of the CIAO software provided by the Chandra X-ray Center (CXC), XSPEC \citep[version 12.7.1;][]{1996ASPC..101...17A}, and Astropy, a community-developed core Python package for Astronomy \citep{2013A&A...558A..33A}.


\begin{thebibliography}{}
\bibitem[Akashi et al.(2008)]{2008NewA...13..563A} Akashi, M., Meiron, Y., \& Soker, N.\ 2008, \na, 13, 563 

\bibitem[Anders \& Grevesse(1989)]{1989GeCoA..53..197A} Anders, E., \& Grevesse, N.\ 1989, \gca, 53, 197 

\bibitem[Arnaud(1996)]{1996ASPC..101...17A} Arnaud, K.~A.\ 1996, Astronomical Data Analysis Software and Systems V, 101, 17 

\bibitem[Astropy Collaboration et al.(2013)]{2013A&A...558A..33A} Astropy Collaboration, Robitaille, T.~P., Tollerud, E.~J., et al.\ 2013, \aap, 558, A33 

\bibitem[Balick \& Frank(2002)]{2002ARA&A..40..439B} Balick, B., \& Frank, A.\ 2002, \araa, 40, 439 

\bibitem[Chu et al.(2001)]{2001ApJ...553L..69C} Chu, Y.-H., Guerrero, M.~A., Gruendl, R.~A., Williams, R.~M., \& Kaler, J.~B.\ 2001, \apjl, 553, L69 

\bibitem[Corradi \& Schwarz(1995)]{1995A&A...293..871C} Corradi, R.~L.~M., \& Schwarz, H.~E.\ 1995, \aap, 293, 871 

\bibitem[Corradi et al.(2003)]{2003MNRAS.340..417C} Corradi, R.~L.~M., Sch{\"o}nberner, D., Steffen, M., \& Perinotto, M.\ 2003, \mnras, 340, 417 

\bibitem[Corradi et al.(2004)]{2004A&A...417..637C} Corradi, R.~L.~M., S{\'a}nchez-Bl{\'a}zquez, P., Mellema, G., Gianmanco, C., \& Schwarz, H.~E.\ 2004, \aap, 417, 637 

\bibitem[Cox et al.(2002)]{2002A&A...384..603C} Cox, P., Huggins, P.~J., Maillard, J.-P., et al.\ 2002, \aap, 384, 603 

\bibitem[Delgado Inglada et al.(2009)]{2009ApJ...694.1335D} Delgado Inglada, G., Rodr{\'{\i}}guez, M., Mampaso, A., \& Viironen, K.\ 2009, \apj, 694, 1335-1348 

\bibitem[Delgado-Inglada \& Rodr{\'{\i}}guez(2014)]{2014ApJ...784..173D} Delgado-Inglada, G., \& Rodr{\'{\i}}guez, M.\ 2014, \apj, 784, 173 

\bibitem[Delgado-Inglada et al.(2016)]{2016MNRAS.456.3855D} Delgado-Inglada, G., Mesa-Delgado, A., Garc{\'{\i}}a-Rojas, J., Rodr{\'{\i}}guez, M., \& Esteban, C.\ 2016, \mnras, 456, 3855 

\bibitem[De Marco(2009)]{2009PASP..121..316D} De Marco, O.\ 2009, \pasp, 121, 316 

\bibitem[Foster et al.(2012)]{2012ApJ...756..128F} Foster, A.~R., Ji, L., Smith, R.~K., \& Brickhouse, N.~S.\ 2012, \apj, 756, 128 

\bibitem[Freeman et al.(2014)]{2014ApJ...794...99F} Freeman, M., Montez, R., Jr., Kastner, J.~H., et al.\ 2014, \apj, 794, 99 

\bibitem[Georgiev et al.(2008)]{2008ApJ...681..333G} Georgiev, L.~N., Peimbert, M., Hillier, D.~J., et al.\ 2008, \apj, 681, 333 

\bibitem[Georgiev et al.(2006)]{2006ApJ...639..185G} Georgiev, L.~N., Richer, M.~G., Arrieta, A., \& Zhekov, S.~A.\ 2006, \apj, 639, 185 

\bibitem[Gruendl et al.(2006)]{2006ApJ...653..339G} Gruendl, R.~A., Guerrero, M.~A., Chu, Y.-H., \& Williams, R.~M.\ 2006, \apj, 653, 339 

\bibitem[Guerrero et al.(2002)]{2002A&A...387L...1G} Guerrero, M.~A., Gruendl, R.~A., \& Chu, Y.-H.\ 2002, \aap, 387, L1 

\bibitem[Kastner et al.(2012)]{2012AJ....144...58K} Kastner, J.~H., Montez, R., Jr., Balick, B., et al.\ 2012, \aj, 144, 58 

\bibitem[Kastner et al.(2008)]{2008ApJ...672..957K} Kastner, J.~H., Montez, R., Jr., Balick, B., \& De Marco, O.\ 2008, \apj, 672, 957 

\bibitem[Kastner et al.(2002)]{2002ApJ...581.1225K} Kastner, J.~H., Li, J., Vrtilek, S.~D., et al.\ 2002, \apj, 581, 1225 

\bibitem[Kastner et al.(2001)]{2001ApJ...550L.189K} Kastner, J.~H., Vrtilek, S.~D., \& Soker, N.\ 2001, \apjl, 550, L189 

\bibitem[Kellogg et al.(2001)]{2001ApJ...563L.151K} Kellogg, E., Pedelty, J.~A., \& Lyon, R.~G.\ 2001, \apjl, 563, L151 

\bibitem[Kellogg et al.(2007)]{2007ApJ...664.1079K} Kellogg, E., Anderson, C., Korreck, K., et al.\ 2007, \apj, 664, 1079 

\bibitem[Kwok et al.(1978)]{1978ApJ...219L.125K} Kwok, S., Purton, C.~R., \& Fitzgerald, P.~M.\ 1978, \apjl, 219, L125 

\bibitem[Latter et al.(2000)]{2000ApJ...539..783L} Latter, W.~B., Dayal, A., Bieging, J.~H., et al.\ 2000, \apj, 539, 783 

\bibitem[Lau et al.(2016)]{2016ApJ...833..115L} Lau, R.~M., Werner, M., Sahai, R., \& Ressler, M.~E.\ 2016, \apj, 833, 115 

\bibitem[Li et al.(2003)]{2003ApJ...590..586L} Li, J., Kastner, J.~H., Prigozhin, G.~Y., \& Schulz, N.~S.\ 2003, \apj, 590, 586 

\bibitem[Li et al.(2004)]{2004ApJ...610.1204L} Li, J., Kastner, J.~H., Prigozhin, G.~Y., et al.\ 2004, \apj, 610, 1204 

\bibitem[Maness et al.(2003)]{2003ApJ...589..439M} Maness, H.~L., Vrtilek, S.~D., Kastner, J.~H., \& Soker, N.\ 2003, \apj, 589, 439 

\bibitem[Marshall et al.(2004)]{2004SPIE.5165..497M} Marshall, H.~L., Tennant, A., Grant, C.~E., et al.\ 2004, \procspie, 5165, 497 

\bibitem[Masson(1989)]{1989ApJ...336..294M} Masson, C.~R.\ 1989, \apj, 336, 294 

\bibitem[Middlemass(1990)]{1990MNRAS.244..294M} Middlemass, D.\ 1990, \mnras,244, 294 

\bibitem[Montez et al.(2005)]{2005ApJ...635..381M} Montez, R., Jr., Kastner, J.~H., De Marco, O., \& Soker, N.\ 2005, \apj, 635, 381 

\bibitem[Montez et al.(2015)]{2015ApJ...800....8M} Montez, R., Jr., Kastner, J.~H., Balick, B., et al.\ 2015, \apj, 800, 8 

\bibitem[Morrison \& McCammon(1983)]{1983ApJ...270..119M} Morrison, R., \& McCammon, D.\ 1983, \apj, 270, 119 

\bibitem[Predehl \& Schmitt(1995)]{1995A&A...293..889P} Predehl, P., \& Schmitt, J.~H.~M.~M.\ 1995, \aap, 293, 889 

\bibitem[Ruiz et al.(2013)]{2013ApJ...767...35R} Ruiz, N., Chu, Y.-H., Gruendl, R.~A., et al.\ 2013, \apj, 767, 35 

\bibitem[Smith et al.(2001)]{2001ApJ...556L..91S} Smith, R.~K., Brickhouse, N.~S., Liedahl, D.~A., \& Raymond, J.~C.\ 2001, \apjl, 556, L91 

\bibitem[Soker \& Kastner(2003)]{2003ApJ...583..368S} Soker, N., \& Kastner, J.~H.\ 2003, \apj, 583, 368 

\bibitem[Soker \& Rappaport(2000)]{2000ApJ...538..241S} Soker, N., \& Rappaport, S.\ 2000, \apj, 538, 241 

\bibitem[Steffen et al.(2008)]{2008A&A...489..173S} Steffen, M., Sch{\"o}nberner, D., \& Warmuth, A.\ 2008, \aap, 489, 173 

\bibitem[Toal{\'a} \& Arthur(2014)]{2014MNRAS.443.3486T} Toal{\'a}, J.~A., \& Arthur, S.~J.\ 2014, \mnras, 443, 3486 

\bibitem[Toal{\'a} \& Arthur(2016)]{2016MNRAS.463.4438T} Toal{\'a}, J.~A., \& Arthur, S.~J.\ 2016, \mnras, 463, 4438 

\bibitem[Volk \& Kwok(1985)]{1985A&A...153...79V} Volk, K., \& Kwok, S.\ 1985, \aap, 153, 79 

\bibitem[Walton et al.(1988)]{1988A&A...200L..21W} Walton, N.~A., Pottasch, S.~R., Reay, N.~K., \& Taylor, A.~R.\ 1988, \aap, 200, L21 

\bibitem[Yu et al.(2009)]{2009ApJ...690..440Y} Yu, Y.~S., Nordon, R., Kastner, J.~H., et al.\ 2009, \apj, 690, 440 

\end{thebibliography}
\end{document}